\documentclass{cimento}
\title{Chasing Lambda}
\author{A.~Kurek\from{ins:x} \atque M.~Szyd{\l}owski\from{ins:y}\from{ins:z}
\instlist{\inst{ins:x}Astronomical Observatory, Jagiellonian University,
Orla 171, 30-244 Krak{\'o}w, Poland
 \inst{ins:y} Mark Kac Complex Systems Research Centre, Jagiellonian University,
Reymonta 4, 30-059 Krak{\'o}w, Poland
\inst{ins:z} Department of Theoretical Physics,
Catholic University of Lublin, Al. Rac{\l}awickie 14, 20-950 Lublin, Poland}}
\shortauthor {Kurek and Szyd{\l}owski}
\PACSes{\PACSit{95.36.+x}{Dark energy} \PACSit{98.80.Es}{Observational cosmology}}
\begin{document}
\maketitle
\begin{abstract}
Recent astronomical observations of SNIa, CMB, as well as BAO in the Sloan Digital Sky Survey, suggest that the current Universe has entered a stage of an accelerated expansion with the transition redshift at $z \simeq 0.5$. While the simplest candidates to explain this fact is cosmological constant/vacuum energy, there exists a serious problem of coincidence. In theoretical cosmology we can find many possible approaches alleviating this problem by applying new physics or other conception of dark energy. We consider state of art candidates for the description of accelerating Universe in the framework of the Bayesian model selection. We point out advantages as well as troubles of this approach. We find that the combination of four data bases gives a stringent posterior probability of the $\Lambda$CDM model which is $74\%$. This fact is a quantitative exemplification of a turmoil in modern cosmology over the $\Lambda$ problem.
\end{abstract}
\section{Introduction}
Recent observations of SNIa stars indicate that the Universe is in an accelerating phase of expansion. If we assume that the FRW model with a source in the form of a perfect fluid well describes present evolution of the Universe, then there is only way to explain the observational fact that the Universe is accelerating, i.e. postulating some gravity source of unknown nature which violates the strong energy condition $\rho+3p>0$. The most natural and simplest candidate for such a type of dark energy is the vacuum energy represented by the cosmological constant $\Lambda$ in the $\Lambda$CDM model. But we have still a problem with this model: predicted value of the vacuum energy density is larger than what we observe by a factor of the order $10^{120}$ (the so called fine tuning problem) and energy density parameters of both dark matter and dark energy are comparable at the present epoch (the so called coincidence problem). Due to that we are still looking for other alternative explanations of accelerating expansion phenomenon. Apart from the dark energy models we have a second group of models where the modification or generalisation of the FRW equations are postulated. We gather the most popular models belonging to the sets described before and we try to answer the question if they are better than the $\Lambda$CDM model in the light of available data sets using the Bayesian model comparison method. 
\section{Cosmological models of the accelerating Universe}
 We consider five models belonging to the group of models with dark energy and gather them together with Hubble functions, with the assumption that the Universe is spatially flat in Table \ref{tab:1}. Here we include the $\Lambda$CDM model \cite{Weinberg:1989}, the phantom dark energy model \cite{Caldwell:2002, Dabrowski:2003}, where the coefficient of the equation of state is a parameter ($w=const$) with the prior assumption that it is less than $-1$, a model with dynamical equation of state \cite{Chevallier:2001,Linder:2003} where the coefficient of the equation of state is a linear function of $a$: $w(a)=w_0+w_1(1-a)$. We also consider the quintessence model where dark energy is represented by a minimally coupled scalar field \cite{Peebles:1988,Ratra:1988}. We choose the power-law parameterized quintessence model, where the evolution of the dark energy density is given by $\rho_X=\rho_{X,0}a^{-3(1+\bar{w}_{X}(a))}$ and $\bar{w}_{X}(a)$ is the mean of a coefficient of the equation of state in the logarithmic scale factor $\bar{w}_{X}(a)=\int w_X(a)\drm(lna)/\int \drm(lna)$ with parametrization proposed by \cite{Rahvar:2007} : $\bar{w}_{X}(a)\equiv w_0 a^{\alpha}$. Finally we include model with the generalized Chaplygin gas \cite{Kamenshchik:2001, Bilic:2002, Bento:2002}, i.e. with perfect fluid which equation of state has the following form $p_X=-A/\rho_X^{\alpha}$, where $A>0$ and $\alpha$ is a constant.
  
\begin{table} 
\caption{The Hubble function for cosmological models with dark energy}
\label{tab:1}
 \begin{tabular}{|r|l|l|}
\hline
\smaller case& \smaller model& \smaller $\frac{H^{2}(z)}{H_0^2}=$\\ 
\hline
&&\\
\smaller 1 &\smaller $\Lambda$CDM model&
\smaller $=\Omega_{\mathrm{m},0}(1+z)^{3}+(1- \Omega_{\mathrm{m},0})$\\ 
\smaller 2& \smaller Model with generalized &
\smaller $=\Omega_{\mathrm{m},0}(1+z)^{3} + (1-\Omega_{\mathrm{m},0})[A_{S}+(1-A_{S})
 (1+z)^{3(1+\alpha)}]^{\frac{1}{1+\alpha}}$ \\
 &\smaller Chaplygin gas & \\
\smaller 3& \smaller Model with phantom & 
\smaller $=\Omega_{\mathrm{m},0}(1+z)^{3}+ 
 (1-\Omega_{\mathrm{m},0})(1+z)^{3(1+w_{X})}$ \\
 &\smaller dark energy & \\
\smaller 4& \smaller Model with dynamical&
\smaller $=\Omega_{\mathrm{m},0}(1+z)^{3}+ 
 (1-\Omega_{\mathrm{m},0})(1+z)^{3(w_{0}+w_{1}+1)} \exp [- \frac {3w_{1}z}{1+z}] $ \\
 &\smaller E.Q.S & \\
\smaller 5&\smaller Quintessence model& 
\smaller $=\Omega_{\mathrm{m},0}(1+z)^{3}+ 
 (1-\Omega_{\mathrm{m},0})(1+z)^{3(1+w_0(1+z)^{-\alpha})}$ \\
\hline
 \end{tabular}
 \end{table}
We also consider five models which belong to the the group of models with the modified or generalized FRW equation (gathered together with the Hubble functions in Table \ref{tab:2}). Here we include examples of so called brane models, which postulated that the observer is embedded on the brane in a larger space in which gravity can propagate: the Dvali-Gabadadze-Porrati model (DGP) \cite{Dvali:2000}, the Sahni-Shtanov brane 1 model \cite{Shtanov:2000, Sahni:2003}. There is also the Cardassian model where the modification of the Friedmann first integral is postulated: $3H^2=\rho+B\rho ^{n}$, and $B$ is a constant \cite{Freese:2002}. We also consider the bouncing model arising in the context of loop quantum gravity (the B$\Lambda$CDM model) \cite{Singh:2005, Szydlowski:2005} and the model with energy transfer between the dark matter and dark energy sectors \cite{Szydlowski:2006}.
\begin{table}
  \caption{The Hubble function for cosmological models with modified theory of gravity.}
  \label{tab:2}
  \begin{tabular}{l|l|l|}
\hline
\smaller case& \smaller model&\smaller $\frac{H^{2}(z)}{H_0^2}=$\\
\hline 
\smaller 6&\smaller DGP&
 \smaller $=\left [ \sqrt{\Omega_{\mathrm{m},0}(1+z)^{3}+\Omega_{rc,0}}+
 \sqrt{\Omega_{rc,0}} \right] ^{2}$ \\    
 & & \smaller $\Omega_{rc,0}=\frac{(1-\Omega_{\mathrm{m},0})^2}{4}$ \\
 & & \\
\smaller 7&\smaller  B$\Lambda$CDM &
 \smaller $=\Omega_{\mathrm{m},0}(1+z)^{3}- 
 \Omega_{n,0}(1+z)^{n}+1-\Omega_{\mathrm{m},0}+\Omega_{n,0}$ \\
& & \\
\smaller 8& \smaller interacting &
 \smaller $=\Omega_{\mathrm{m},0}(1+z)^{3}+\Omega_{\mathrm{int},0}(1+z)^{n}+ 
  1-\Omega_{\mathrm{m},0}-\Omega_{\mathrm{int},0}$ \\
& \smaller with $\Lambda$ & \\
\smaller 9&\smaller Cardassian &
 \smaller $=\Omega_{r,0}(1+z)^{4}+\Omega_{\mathrm{m},0}(1+z)^{4} 
 \left[ \frac{1}{1+z} + (1+z)^{-4+4n} \left ( \frac{1- \Omega_{r,0}-
 \Omega_{\mathrm{m},0}}{\Omega_{\mathrm{m},0}} \right)E(z) \right ]$ \\
 &\smaller $\Omega_{r,0}=10^{-4}$ & \smaller $E(z)=\left ( \frac{ \frac{1}{1+z}+
 \frac{\Omega_{r,0}}{\Omega_{\mathrm{m},0} }}{ 1+ 
 \frac{\Omega_{r,0}}{\Omega_{\mathrm{m},0}}}\right)^{n}$ \\
& & \\
\smaller 10& \smaller Sahni-Shtanov&
 \smaller $=\Omega_{\mathrm{m},0}(1+z)^{3}+\Omega_{\sigma,0}+ 
 2 \Omega_{l,0}-2\sqrt{\Omega_{l,0}}P(z)$ \\
&\smaller brane I & 
\smaller $\Omega_{\sigma,0}=1-\Omega_{\mathrm{m},0}+2\sqrt{\Omega_{l,0}}\sqrt{1+\Omega_{\Lambda b,0}}$ \\
& & \smaller $P(z)=\sqrt{\Omega_{\mathrm{m},0}(1+z)^{3}+
 \Omega_{\sigma,0}+ \Omega_{l,0}+ \Omega_{\Lambda b,0}}$ \\
\hline
  \end{tabular}
\end{table}
\section{Bayesian model comparison framework and application to cosmological models}
In the Bayes theory the best model from the set under consideration is the one which has the largest value of the posterior probability \cite{Jeffreys:1961} (the probability in the light of data $D$) which has the following form
\begin{equation}
 P(M|D)=\frac{P(D|M)P(M)}{P(D)}.
\end{equation}
$P(M)$ is the prior probability for the model considered, the value of which depends on our previous knowledge (i.e. without information coming from the data $D$) about the model. If we have no foundation to favour one model over another one from the set of models under consideration we usually assume the same value of the prior quantity for all models, i.e. $P(M)=1/K$, where $K$ is the number of models. $P(D)$ is the normalization constant. Requiring that a sum of posterior probabilities for all models from the considered set is equal to one we can obtain the form for this quantity:
\begin{equation}
 \sum _{i=1}^{K} P(M_i|D)=1 \  \longrightarrow P(D)=\sum_{i=1}^{K}P(D|M_i)P(M_i),
\end{equation}
where $i$ indexed the model under consideration. It should be pointed out that conclusions based on the posterior probability depend on the set of models under consideration and could change if the set is changed.

$P(D|M)$, the marginal likelihood (also called the evidence), is the most important quantity in the Bayesian framework of model comparison and has the following form
\begin{equation}
P(D|M)=\int L(\bar{\theta}|D,M)P(\bar{\theta}|M)\drm\bar{\theta} \equiv E,
\end{equation}
where $L(\bar{\theta}|D,M)$ is the likelihood of the model $M$, $\bar{\theta}$ is the vector of the model parameters and $P(\bar{\theta}|M)$ is the prior probability for the model parameters. Here we use an asymptotic approximation to the minus twice logarithm of the evidence derived by Schwarz \cite{Schwarz:1978}, so called BIC quantity which has the following form:
\begin{equation}
 \mathrm{BIC}=-2 \ln \mathcal{L} + d \ln N,
\end{equation}  
where $\mathcal{L}$ is the maximum of the likelihood function, $d$ is the number of model parameters and $N$ is the number of data used in the analysis. This approximation is derived with the assumption that observations ($D=\{x_i\}, i=1,\cdots,N$) used in analysis are iid (independent and identically distributed) and comes from a linear exponential distribution defined in the following way $ f(x_{i}|\bar{\theta})= \exp \left [\sum _{k=1}^{S}w_{k}(\bar{\theta})t_{k}(x_{i})+b(\bar{\theta})\right ]$, where $w_{1},\dots,w_{S},b$ are functions of only $\bar{\theta}\in \mathbf{R}^{d}$,  $t_{1},\dots,t_{S}$ are function of only $x_{i}$ and $S=d$. The prior for model parameters  must be not equal to $0$ in the point where the likelihood function reach maximum and the number of data must be large enough with respect to the number of model parameters (as a consequence of asymptotic approximation). This derivation was generalized by \cite{Cavanaugh:1999} where it is not required to assume any specific form for the likelihood function but is only necessary that the likelihood function satisfies some non-restrictive regularity conditions. Moreover the data do not need to be iid. This derivation requires to assume that a prior for model parameters is not equal to zero in the neighborhood of the point where the likelihood function under a given model reaches a maximum and that it is bound in the whole parameter space under consideration. The number of the data must be also large with respect to the number of model parameters.

We apply this method to compare the cosmological models described before. We make the comparison among the models with dark energy (gathered in Table \ref{tab:1}), among the models with modified theory of gravity (gathered in Table \ref{tab:2}) as well as among all models described before. We use the information coming from the SNIa, CMB, BAO and H(z) data. Let us describe those data sets.

We use $N_1=192$ SNIa data \cite{Riess:2007, Wood:2007, Davis:2007}. In this case the likelihood function has the following form
\begin{equation}
 L_{SN}\propto \exp \left[-\frac{1}{2}\left(\sum_{i=1}^{N_1}\frac{(\mu_{i}^{\mathrm{theor}}-\mu_{i}^{\mathrm{obs}})^{2}}{\sigma_{i}^{2}}\right) \right],
\end{equation}
where $\sigma_{i}$ is known, $\mu_{i}^{\mathrm{obs}}=m_{i}-M$ ($m_{i}$--apparent magnitude, $M$--absolute magnitude of SNIa), $\mu_{i}^{\mathrm{theor}}=5\log_{10}D_{Li} + \mathcal{M}$, $\mathcal{M}=-5\log_{10}H_{0}+25$ and $D_{Li}=H_{0}d_{Li}$, where $d_{Li}$ is the luminosity distance, which with assumption that $k=0$ is given by $d_{Li}=(1+z_{i})c\int_{0}^{z_{i}}(\drm z'/H(z'))$.

We also include information coming from the CMB data. Here the likelihood function has the following form
\begin{equation}
L_{R} \propto \exp \left[- \frac {(R^{\mathrm{theor}}-R^{\mathrm{obs}})^2}{2\sigma_{R}^2} \right],
\end{equation}
where $R$ is the shift parameter, $R^{\mathrm{theor}}=\sqrt{\Omega_{\mathrm{m},0}}\int_{0}^{z_{\mathrm{dec}}}(H_0/H(z))\drm z$, and $R^{\mathrm{obs}}=1.70 \pm 0.03$ for $z_{\mathrm{dec}}=1089$ \cite{Spergel:2006, Wang:2006ts}.

As third observational data we use the measurement of the baryon acoustic oscillations (BAO) from the SDSS luminous red galaxies \cite{Eisenstein:2005}. In this case the likelihood function has the following form
\begin{equation}
 L_{A} \propto \exp \left[ -\frac{(A^{\mathrm{theor}}-A^{\mathrm{obs}})^2}{2\sigma_{A}^2}\right ],
\end{equation}
where $A^{\mathrm{theor}}=\sqrt{\Omega_{\mathrm{m},0}} \left (H(z_A)/H_{0} \right ) ^{-1/3} \left [(1/z_{A}) \int_{0}^{z_{A}}(H_0/H(z))\drm z\right]^{2/3}$ and $A^{\mathrm{obs}}=0.469 \pm 0.017$ for $z_{A}=0.35$.

Finally we use the observational $H(z)$ data ($N_2=9$) from \cite{Simon:2005} (see also \cite{Samushia:2006, Wei:2007} and references therein). These data are based on the differential ages ($\drm t/\drm z$) of the passively evolving galaxies which allow to estimate the relation $H(z) \equiv \dot{a}/a =-1/(1+z) (\drm z/\drm t)$. Here the likelihood function has the following form
\begin{displaymath}
 L_{H} \propto \exp \left(-\frac{1}{2} \left[ \sum_{i=1}^{N_2}\frac{\left( H(z_i) -H_i(z_i) \right) ^2}{\sigma_i ^2} \right] \right),
\end{displaymath}
where $H(z)$ is the Hubble function, $H_i,\ z_i$ are observational data.

The final likelihood function is given by $L=L_{SN}L_{R}L_{A}L_{H}$, with the number of data points $N=192+1+1+9$. We assume equal prior probabilities for all models from the set considered. Results, i.e. the values of the posterior probabilities, for the set of models with dark energy and set of models with modified theory of gravity are gathered in Table \ref{tab:3} while for the set of all models in Table \ref{tab:4}.

\begin{table}
\caption{Posterior probabilities for models from Table~\ref{tab:1} and for models from Table~\ref{tab:2}.}
\label{tab:3}
\begin{narrowtabular}{1cm}{ccc||ccc}
model & prior & posterior & model&prior&posterior \\
\hline
1 & 0.20 & \textbf{0.84}& 6 & 0.20 & 0.07\\
2 & 0.20 & 0.02& 7 & 0.20 & 0.03\\
3 & 0.20 & 0.06 & 8 & 0.20 & 0.13\\
4 & 0.20 & 0.04 & 9 & 0.20 & \textbf{0.74}\\
5 & 0.20 & 0.04 & 10 & 0.20 & 0.03\\
\hline
\end{narrowtabular}
\end{table}
\begin{table}
\caption{Posterior probabilities for models from Tables~\ref{tab:1} and \ref{tab:2}.}
\label{tab:4}
\begin{narrowtabular}{2cm}{ccc}
model&prior&posterior \\
\hline
1 & 0.10 & \textbf{0.74}\\
2 & 0.10 & 0.02\\
3 & 0.10 & 0.05\\
4 & 0.10 & 0.04\\
5 & 0.10 & 0.03\\
6 & 0.10 & 0.01\\
7 & 0.10 & 0.005\\
8 & 0.10 & 0.01\\
9 & 0.10 & 0.09\\
10 & 0.10 & 0.005\\
\hline
\end{narrowtabular}
\end{table}
We can conclude that the $\Lambda$CDM model is the best one in the set of model with dark energy as well as the best one among all considered model. If we compare only models with modified theory of gravity the best one from them is the Cardassian model.
\section{Conclusion}
Our results of Bayesian method of model selection indicate that the $\Lambda$CDM is the best one. On the other hand this model has the status of an effective theory which gives us a good description, however without deep understanding. The $\Lambda$CDM model is based on the general relativity theory which is believed to be superposed by
a more fundamental theory, namely quantum gravity theory. Therefore,
we still search for a quantum based model of the Universe
\cite{Stachowiak:2006uh,Mielczarek:2007zy}. So far the $\Lambda$CDM
model is some effective theory of the Universe the best we have.
In our opinion the $\Lambda$CDM model is expected to be emergent from
the future quantum gravity theory of the Universe and finding it is
the challenge of the 21st century.

\acknowledgments
This work has been supported by the Marie Curie Actions Transfer of Knowledge project
COCOS (contract MTKD-CT-2004-517186) and by the project Particle Physics and Cosmology (contract MTKD-CT-2005-029466). The authors are very grateful to T. Stachowiak for helpful discussion and suggestions.

\end{document}